\documentstyle[12pt,epsf,rotate,epic,eepic]{article} 
\if@twoside \oddsidemargin 21pt \evensidemargin 59pt \marginparwidth 85pt
\else \oddsidemargin 0pt \evensidemargin 0pt
\fi
\newcounter{saveeqn}                  
\newcommand{\alpheqn}[1]{\refstepcounter{equation}\label{#1}%
\setcounter{saveeqn}{\value{equation}}%
\setcounter{equation}{0}%
\renewcommand{\theequation}                  
{\mbox{\arabic{saveeqn}\alph{equation}}}}
\newcommand{\reseteqn}{\setcounter{equation}{\value{saveeqn}}%
\renewcommand{\theequation}{\arabic{equation}}}
\newcounter{savefig}

\newcommand{\capt}[1]{\newcommand{\groesse}{\normalsize}%
\renewcommand{\normalsize}{\small}%
\caption[ ]{#1}%
\renewcommand{\normalsize}{\groesse}}
\renewcommand{\vec}[1]{\mbox{\boldmath $#1$}}
\newcommand{\vek}[1]{\mbox{\boldmath $#1$}}
\newcommand{\weg}[1]{#1}                       
\begin{document}
\title{Stochastic and {\sc Boltzmann}-like
models for behavioral changes, and their relation to
game theory}

\author{Dirk Helbing} 
\maketitle

\begin{abstract}
\mbox{ }\\[-0.5cm]
In the last decade, stochastic models have shown to be very useful for
quantitative modelling of social processes.
Here, a configurational master equation
for the description of behavioral changes by pair
interactions of individuals is developed. 
Three kinds of social
pair interactions are distinguished: Avoidance processes, compromising
processes, and imitative processes.  
Computational results are presented for a special case of imitative
processes: the competition of two equivalent strategies.
They show a phase transition that describes the selforganization of a
behavioral convention.
This phase transition is further analyzed by examining the equations for
the most probable behavioral distribution, which are
{\sc Boltzmann}-like equations. Special cases of {\sc Boltzmann}-like
equations do not obey the $H$-theorem and have oscillatory or even
chaotic solutions.
A suitable {\sc Taylor} approximation leads
to the socalled game dynamical equations (also known as
selection-mutation equations in the theory of evolution). 
\end{abstract}

\section{Introduction}

It is well-known that {\sc Markov}ian stochastic processes can be described
by a {\em master equation}. 
The master equation has found many applications in thermodynamics 
\cite{Zwan}, chemical kinetics \cite{Opp}, 
laser theory \cite{Hak} and biology \cite{Arn}. 
Moreover, in the last decade {\sc Weid\-lich} and
{\sc Haag} have successfully introduced it for the description of 
social processes \cite{WeHa83,We91} like opinion formation \cite{Weid1}, 
migration \cite{Weid2}, agglomeration \cite{WeHa87} and
settlement processes \cite{WeMu90}. 
\par
Since the master equation is difficult to solve (even numerically) one
often examines the equations for the most probable distribution of states,
instead. These equations are found to be 
``{\sc Boltzmann}-{\em like}'' {\em equations},
and have many applications to the kinetics of
gases \cite{Bo64} or chemical reactions 
\cite{Br82}. Special cases of 
{\sc Boltzmann}-like equations have also become increasingly important
in quantitative social science, namely
the {\em logistic equation} for the description of limited
growth processes \cite{Pea24,Ve45} and
the socalled {\em gravity model} for intercity migration processes
\cite{Zi46}.
Recently, {\sc Boltzmann}-like models have been 
suggested for avoidance processes of 
pedestrians \cite{Diss,Compl}, and for attitude formation by direct pair
interactions of individuals occuring in discussions \cite{Diss,He92}. 
The models for attitude formation include cases of 
{\em oscillatory} or even {\em chaotic}
behavior (see sect. \ref{CHaoS}). Such behavior is, for example, known from
fashion or economics (economic cycles, stock market).
\par
In the following, a master equation for behavioral changes by
spontaneous transitions and pair interactions will be developed.
It allows the description of the selforganization of behavioral conventions.
Three kinds of social pair interactions are distinguished: Imitative processes,
avoidance processes, and compromising processes. It turns out 
that for a special case of imitative processes the 
{\em game dynamical equations} result, which are used 
for the description of cooperation and competition processes.     
\par
The game dynamical equations are empiricaly validated
\cite{Co84,Ra90}, and have many important applications in
social sciences \cite{Ax84,Mue90} and economy \cite{NeuMo44}. They are also a 
powerful tool in 
evolutionary biology \cite{Ei71,Fi30,EiSchu79,FeEb89}.
Moreover, the {\sc Lotka-Volterra}-{\em equations} \cite{Lo56,Vo31}
for the description of predator-prey systems in ecology 
\cite{GoMaMo71} are mathematically
equivalent to a special class of game dynamical equations \cite{HoSi88}.

\section{The configurational master equation}

Suppose we have a system 
with a large number $N \gg 1$ of subsystems
(e.g. a gas with $N$ atoms). These subsystems are distributed over 
several states $\vec{x}$ (which e.g. distinguish the places $\vec{r}$
and velocities $\vec{v}$ of the atoms). If the {\em occupation number}
$n_{\weg{x}}$ means the number of subsystems that are in state $\vec{x}$,
we have the relation
\begin{equation}
 \sum_{\weg{x}} n_{\weg{x}} = N \, .
\end{equation}
The vector
\begin{equation}
 \vec{n} := (\dots,n_{\weg{x}},\dots)^{\rm tr}
\end{equation}
consisting of the
occupation numbers is called the {\em configuration} of the system
(since it contains all information about the distribution of the $N$
subsystems over the states $\vec{x}$). 
$P(\vec{n},t)$ shall denote the
probability to find the configuration $\vec{n}$ at time $t$. This implies
\begin{equation}
 0 \le P(\vec{n},t) \le 1  \qquad \mbox{and} \qquad 
\sum_{\weg{n}} P(\vec{n},t) = 1 \, .
\end{equation}
The temporal development of the probability $P(\vec{n},t)$ 
is governed by a {\em master equation} \cite{Helb}:
\begin{eqnarray}
 \frac{d}{dt} P(\vec{n},t) &=&
\mbox{inflow into $\vec{n}$ } \qquad \quad 
- \mbox{ outflow from $\vec{n}$} \nonumber \\
&=& \sum_{\weg{n}'} w(\vec{n}|\vec{n}';t)P(\vec{n}',t) 
- \sum_{\weg{n}'} w(\vec{n}'|\vec{n};t)P(\vec{n},t) \, . \qquad
\label{master}
\end{eqnarray}
$w(\vec{n}'|\vec{n};t)$ are the {\em configurational transition rates}
of transitions from configuration $\vec{n}$ to configuration $\vec{n}'$.
Often the dynamics of the system is mainly given by 
{\em spontaneous transitions} and {\em direct pair interactions} of 
subsystems. In this case, the configurational transition rates are of the
following form \cite{Helb}
\begin{equation}
w(\vec{n}'|\vec{n};t) := \left\{
\begin{array}{ll}
w_1(\vec{x}'|\vec{x};t) n_{\weg{x}} 
& \mbox{if } \vec{n}' = \vec{n}_{\weg{x}'\weg{x}} \\
w_2(\vec{x}',\vec{y}'|\vec{x},\vec{y};t) n_{\weg{x}} n_{\weg{y}} 
& \mbox{if } \vec{n}' = \vec{n}_{\weg{x}'\weg{y}'\weg{x}\weg{y}} \\
0 & \mbox{otherwise.}
\end{array}\right. 
\label{rate}
\end{equation}
\begin{itemize}
\item Spontaneous changes of the state from $\vec{x}$ to $\vec{x}'$
with an {\em individual} transition rate 
$w_1(\vec{x}'|\vec{x};t)$ correspond to transitions of the
configuration from $\vec{n}$ to
\begin{equation}
 \vec{n}_{\weg{x}'\weg{x}} := (\dots,(n_{\weg{x}'} + 1)
\dots, (n_{\weg{x}} -1), \dots )^{\rm tr} \,
\end{equation}
with a {\em configurational} transition rate $w(\vec{n}_{\weg{x}'\weg{x}}|
\vec{n};t)=w_1(\vec{x}'|\vec{x};t) n_{\weg{x}}$, which is
proportional to the number $n_{\weg{x}}$ of subsystems that can change the
state $\vec{x}$.
\item Pair interactions leading one subsystem to change the state
from $\vec{x}$ to $\vec{x}'$ and another subsystem to change the state from
$\vec{y}$ to $\vec{y}'$ correspond to a transition of the configuration from
$\vec{n}$ to 
\begin{equation}
 \vec{n}_{\weg{x}'\weg{y}'\weg{x}\weg{y}} 
:= (\dots,(n_{\weg{x}'} + 1),\dots, 
(n_{\weg{x}} - 1),\dots,(n_{\weg{y}'} + 1),
\dots,(n_{\weg{y}} - 1),\dots )^{\rm tr}
\end{equation}
with a configurational transition rate 
$w(\vec{n}_{\weg{x}'\weg{y}'\weg{x}\weg{y}}|\vec{n};t) = 
w_2(\vec{x}',\vec{y}'|\vec{x},\vec{y};t) n_{\weg{x}}n_{\weg{y}}$, which
is proportional to the number $n_{\weg{x}}n_{\weg{y}}$ of possible pair
interactions between subsystems that are in state $\vec{x}$ resp. 
$\vec{y}$ (if $n_{\weg{x}} \gg 1$ where $P(\vec{n},t)$ is not 
negligible, see \cite{Helb}).
\end{itemize}
The description of social processes often requires {\em generalized}
configurational transition rates of the form
\begin{equation}
w(\vec{n}'|\vec{n};t) := \left\{
\begin{array}{ll}
w_1(\vec{x}'|\vec{x};\vec{n};t) n_{\weg{x}} 
& \mbox{if } \vec{n}' = \vec{n}_{\weg{x}'\weg{x}} \\
w_2(\vec{x}',\vec{y}'|\vec{x},
\vec{y};\vec{n};t) n_{\weg{x}} n_{\weg{y}} 
& \mbox{if } \vec{n}' = \vec{n}_{\weg{x}'\weg{y}'\weg{x}\weg{y}} \\
0 & \mbox{otherwise,}
\end{array}\right. 
\end{equation}
since individuals may react on the actual (socio)configuration $\vec{n}$.
The dependence of the {\em individual} transition rates $w_1$ and
$w_2$ on $\vec{n}$ reflects {\em indirect interactions}
of the individuals.

\section{Equations for behavioral changes}\renewcommand{\vec}[1]{#1}

For the description of a system of $N$ individuals, the states $\vec{x} \in
\{1,\dots,S\}$ shall represent the possible {\em behavioral strategies} of
individuals concerning a certain situation. 
The pair interactions
\begin{equation}
 \vec{x}',\vec{y}'\, \longleftarrow \, \vec{x},\vec{y} \, , \qquad
\label{allkla}
\end{equation}
during which the strategies are changed from $\vec{x}$ and $\vec{y}$ to 
$\vec{x}'$ and $\vec{y}'$,
can be completely classified according to the following scheme:
\begin{equation}
\left.
\begin{array}{rcl}
\vec{x},\vec{x} &\longleftarrow & \vec{x},\vec{x} \\
\vec{x},\vec{y} &\longleftarrow & \vec{x},\vec{y}
\end{array} \right\} (0)
\end{equation}
\begin{equation}
\left.
\begin{array}{rcll}
\vec{x},\vec{x} &\longleftarrow & \vec{x},\vec{y} & (\vec{x}\ne \vec{y})\\
\vec{y},\vec{y} &\longleftarrow & \vec{x},\vec{y} & (\vec{x}\ne \vec{y})
\end{array} \right\} ({\rm I})
\end{equation}
\begin{equation}
\left.
\begin{array}{rcll}
\vec{x},\vec{y}' &\longleftarrow & \vec{x},\vec{x} & (\vec{y}'\ne \vec{x})\\
\vec{x}',\vec{y} &\longleftarrow & \vec{y},\vec{y} & (\vec{x}'\ne \vec{y})\\
\vec{x}',\vec{y}'&\longleftarrow & \vec{x},\vec{x} & (\vec{x}'\ne \vec{x},
 \vec{y}'\ne \vec{x})
\end{array} \right\} ({\rm II})
\end{equation}
\begin{equation}
\left.
\begin{array}{rcll}
\vec{x},\vec{y}' &\longleftarrow & \vec{x},\vec{y} & (\vec{x}\ne \vec{y}, 
 \vec{y}'\ne \vec{y},\vec{y}'\ne \vec{x})\\
\vec{x}',\vec{y} &\longleftarrow & \vec{x},\vec{y} & (\vec{x}\ne \vec{y}, 
 \vec{x}'\ne \vec{x}, \vec{x}'\ne \vec{y})\\
\vec{x}',\vec{y}'&\longleftarrow & \vec{x},\vec{y} & (\vec{x}\ne \vec{y}, 
 \vec{x}'\ne \vec{x},\vec{y}'\ne \vec{y}, \vec{x}'\ne \vec{y}, \vec{y}'\ne 
 \vec{x}) 
\end{array}\right\} ({\rm III})
\end{equation}
\begin{equation}
\left.
\begin{array}{rcll}
\vec{y},\vec{x} &\longleftarrow & \vec{x},\vec{y} & (\vec{x}\ne \vec{y})\\
\vec{x}',\vec{x} &\longleftarrow & \vec{x},\vec{y} & (\vec{x}\ne 
 \vec{y},\vec{x}'\ne \vec{x}, \vec{x}'\ne \vec{y})\\
\vec{y},\vec{y}'&\longleftarrow & \vec{x},\vec{y} & (\vec{x}\ne 
 \vec{y},\vec{y}'\ne \vec{y},\vec{y}'\ne \vec{x}) 
\end{array} \right\} ({\rm IV}) 
\end{equation}
Obviously, the interpretation of the above {\em kinds}
$k \in \{0, {\rm I},\dots,{\rm IV} \}$ of pair interactions is the
following:
\begin{itemize}
\item[(0)] During interactions of kind (0) both individuals do not change
their strategy. These interactions can be omitted in the following, 
since they have no contribution to the change of $P(\vek{n},t)$. 
\item[(I)] The interactions (I) describe {\em imitative processes}
(processes of persuasion), i.e., the tendency to take over the strategy of
another individual. 
\item[(II)] The interactions (II) describe {\em avoidance processes}, where an
individual changes the strategy when meeting another individual using the
same strategy. (Processes of this kind are known as aversive behavior, defiant
behavior or snob effect.)
\item[(III)] The interactions (III) represent some 
kind of {\em com\-pro\-mi\-sing
pro\-cesses}, where an individual changes the strategy to a new one
(the ``compromise'') when meeting an individual with another strategy.
(Such processes are found, if a certain strategy cannot be maintained when
confronted with another strategy.)
\item[(IV)] The interactions (IV) 
describe imitative processes, in which an 
individual changes the strategy despite of the fact, that he or she 
convinces the interaction partner of his resp. her strategy. 
Social processes of this
kind are very improbable and can normally be neglected.
\end{itemize}
The different kinds of pair interactions have been discussed in
\cite{Diss,He92,Hel92}. 
In the following, our considerations are restricted to imitative
processes. The corresponding individual transition rates have, then, the
following general form:
\begin{eqnarray}
w_2(\vec{x}',\vec{y}'|\vec{x},\vec{y};\vek{n};t)
&=& w_2^*(\vec{x}|\vec{y};\vek{n};t) \delta_{\weg{x}\weg{x}'}
\delta_{\weg{x}\weg{y}'}(1-\delta_{\weg{x}\weg{y}}) \nonumber \\
&+& w_2^*(\vec{y}|\vec{x};\vek{n};t) \delta_{\weg{y}\weg{y}'}
\delta_{\weg{y}\weg{x}'} (1-\delta_{\weg{x}\weg{y}}) \, . \qquad
\label{imita}
\end{eqnarray}
$w_2^*(\vec{y}|\vec{x};\vek{n};t)$ is 
the rate of imitative strategy changes from
$\vec{x}$ to $\vec{y}$ and shall be specified now:
Let $A_{\weg{x}\weg{x}'}$ be the {\em success} of strategy $\vec{x}$ when 
confronted with strategy $\vec{x}'$. Then, 
\begin{equation}
 E(\vec{x};\vek{n};t) := \sum_{\weg{x}'} A_{\weg{x}\weg{x}'}
 \frac{n_{\weg{x}'}(t)}{N}
\end{equation}
is the {\em expected success} of strategy $\vec{x}$ in interactions with other
strategies. 
With
\begin{equation}
 w_2^*(\vec{y}|\vec{x};\vek{n};t) :=
\frac{\exp \Big[ E(\vec{y};\vek{n};t) - E(\vec{x};\vek{n};t) \Big]}
{D(\vec{y},\vec{x};t)} \, ,
\label{mult}
\end{equation}
imitative strategy changes from $\vec{x}$ to $\vec{y}$
will occur the more frequent, the
greater the expected increase 
\begin{equation}
 \Delta_{\weg{y}\weg{x}} E := E(\vec{y};\vek{n};t) - E(\vec{x};\vek{n};t)
\end{equation}
of success is, and the smaller the {\em incompatibility (``distance'')}
\begin{equation}
 D(\vec{y},\vec{x};t) \equiv D(\vec{x},\vec{y};t) > 0
\end{equation}
between the strategies $\vec{x}$ and $\vec{y}$ is. (\ref{mult}) is a 
variant of the {\em multinomial logit model} 
\cite{Diss,DoFa75}, which has shown to
be suitable for the description of decision processes. The ansatz
(\ref{mult}) can also be derived by entropy maximization \cite{Diss}
or with the {\sc Fechner}ian law of psychophysics \cite{Diss,Lu59}. 

\section{Selforganization of behavioral conventions by competition between
strategies}

As an example for the behavioral equations, we shall consider a case
where the individuals can choose between two
{\em equivalent} strategies $\vec{x} \in \{1,2\}$, 
i.e., the {\em payoff matrix}
$\underline{A}$ shall be symmetrical:\\[-0.5cm]
\begin{equation}
 \underline{A} \equiv \Big( A_{\weg{x}\weg{x}'} \Big) :=
\left(
\begin{array}{cc}                                   
A+B & B \\                                                   
B & A+B
\end{array}\right) \, .                                 
\label{pay}
\end{equation}
For spontaneous strategy changes 
we shall assume the simplest form of transition rates:
\begin{equation}
 w_1(\vec{x}'|\vec{x};\vek{n};t) \equiv W \, .
\label{fluct}
\end{equation}
A situation of the above kind is the avoidance behavior of pedestrians 
\cite{Diss,He91}:
In pedestrian crowds with two opposite directions of movement, the
pedestrians have sometimes to avoid each other in order to exclude a collision.
For an avoidance maneuver to be successful, both pedestrians concerned have to
pass the respective other pedestrian either on the right hand side
($\vec{x}=1$) or on
the left hand side ($\vec{x}=2$). Otherwise, both pedestrians have to stop.
Therefore, both strategies 
(to pass pedestrians on the right hand side 
or to pass them on the left hand side)
are equivalent, but the success of a strategy grows
with the number $n_{\weg{x}}$ of individuals who use the {\em same} strategy.
In the payoff matrix (\ref{pay}) we have $A>0$, then.
\begin{figure}[htbp]
\epsfysize=5.2cm 
\centerline{\hbox{\epsffile[108 575 370 724]{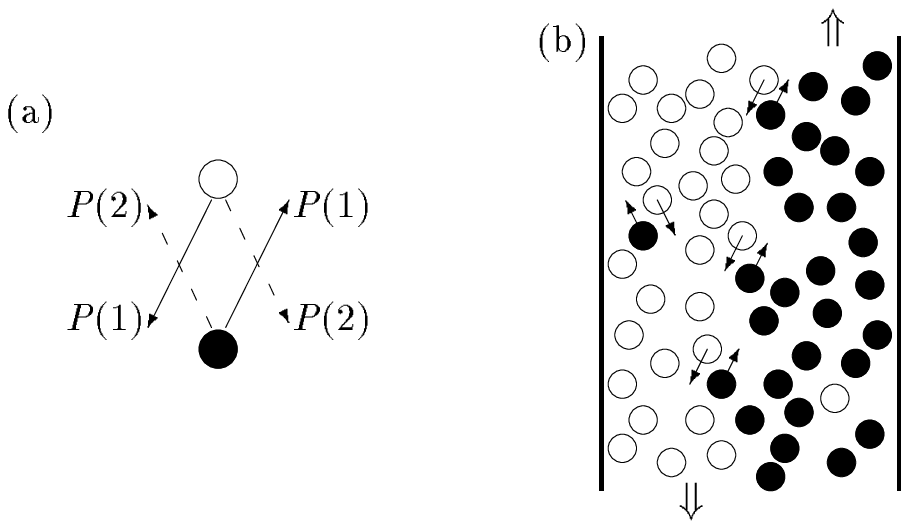}}}
\capt{(a) For pedestrians with an opposite 
direction of motion it is advantageous,
if both prefer either the right hand side or the left hand side when
trying to pass each other. Otherwise, they would have to stop in order
to avoid a collision.\\
(b) Opposite directions of motion normally use separate lanes.
Avoidance maneuvers are indicated by 
arrows.\label{separation}}
\end{figure}
\par
Empirically one finds that 
the probability $P(1)$ for choosing the right hand side
is usually different from the probability $P(2)=1-P(1)$ 
for choosing the left hand side (see fig. \ref{separation}a). 
As a consequence, opposite directions of motion 
normally use separate lanes (see fig. \ref{separation}b).
\par
\begin{figure}[htbp]
\parbox[b]{7.8cm}{                                                   
\epsfxsize=7cm 
\centerline{\rotate[r]{\hbox{\epsffile[28 28 570                              
556]{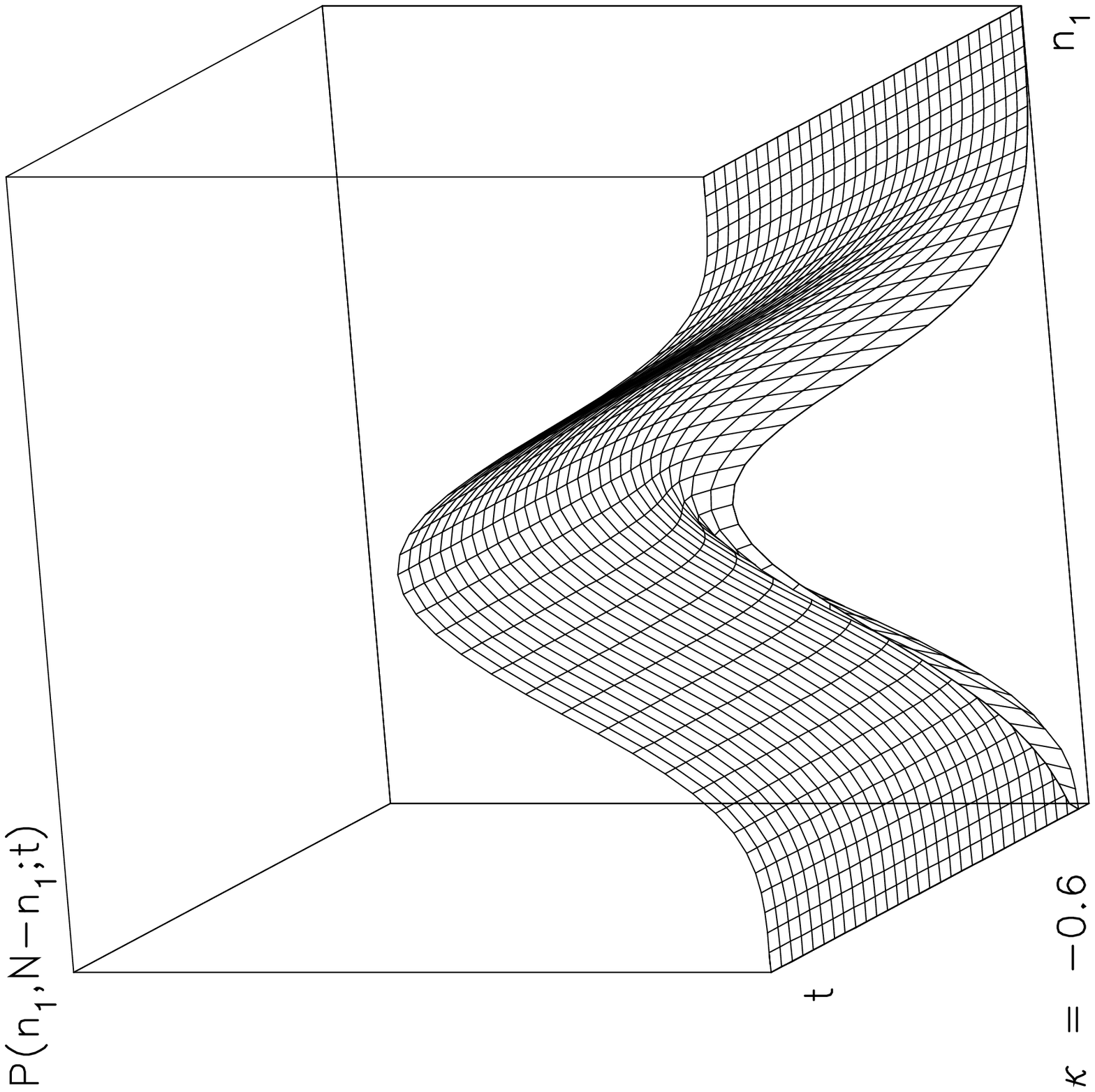}}}}
}\hfill
\parbox[b]{7.8cm}{
\epsfxsize=7cm 
\centerline{\rotate[r]{\hbox{\epsffile[28 28 570
556]{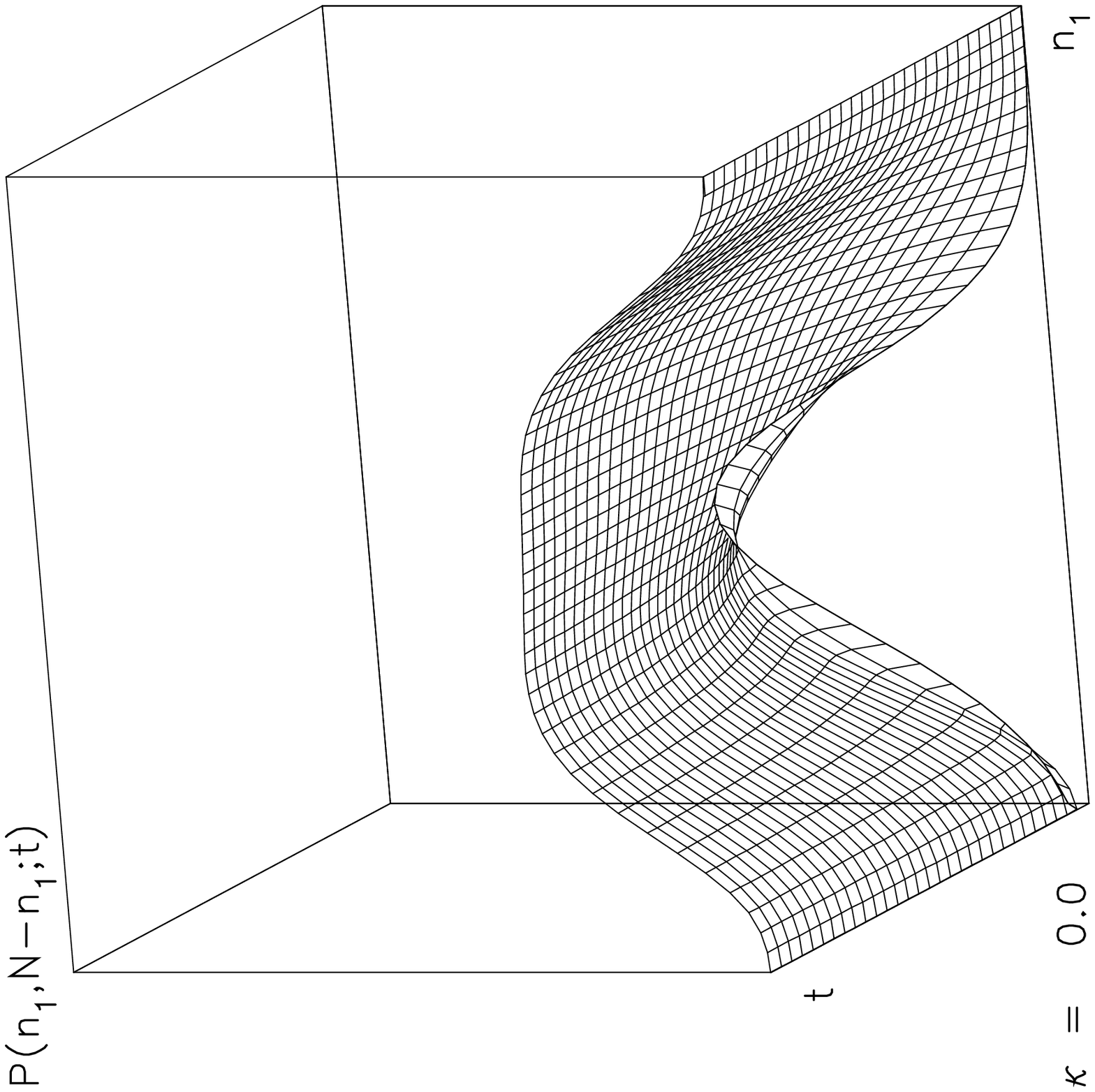}}}}
}                                              
%
\parbox[b]{7.8cm}{
\epsfxsize=7cm 
\centerline{\rotate[r]{\hbox{\epsffile[28 28 570
556]{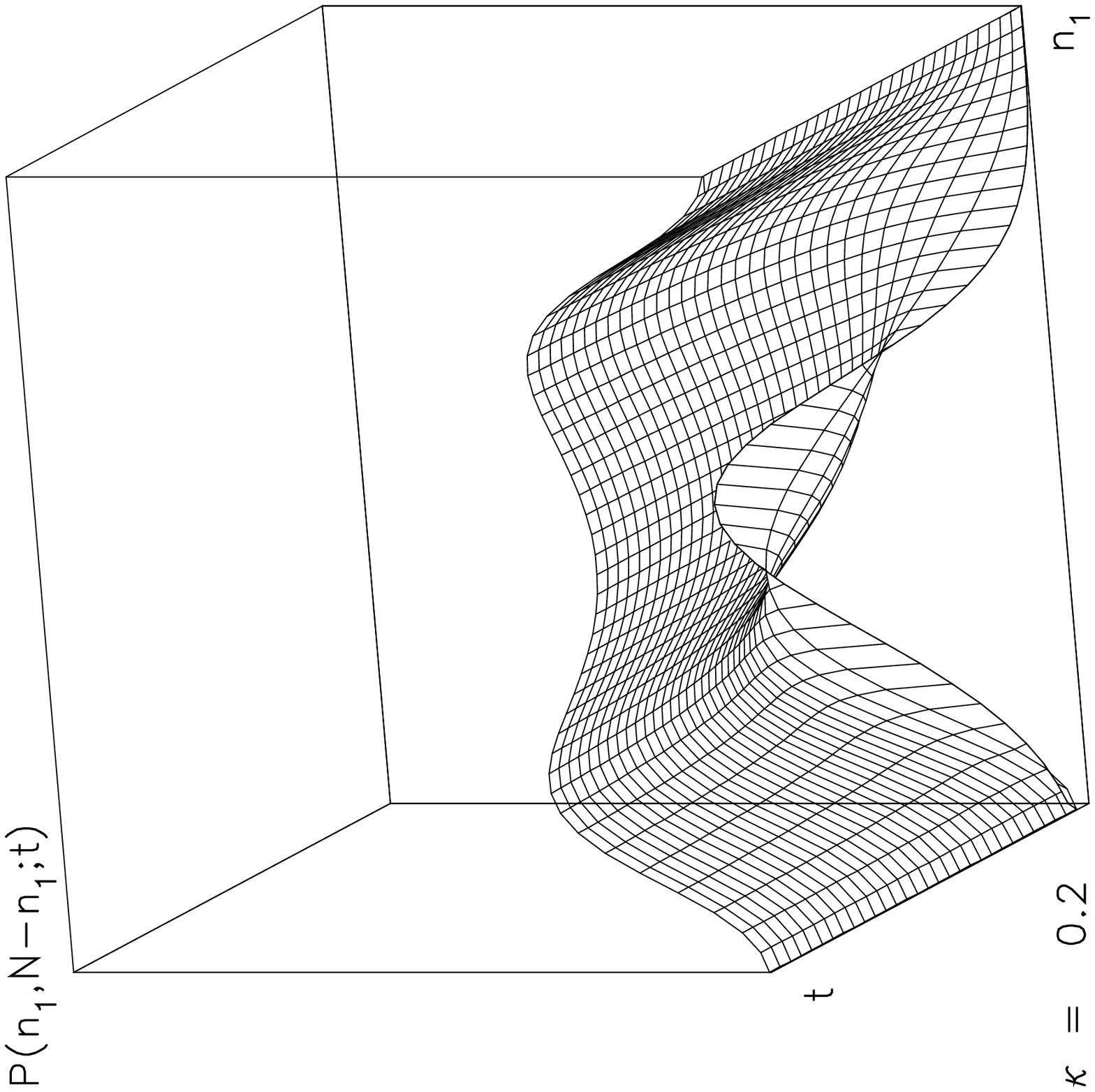}}}}
}\hfill
\parbox[b]{7.8cm}{
\epsfxsize=7cm 
\centerline{\rotate[r]{\hbox{\epsffile[28 28 570
556]{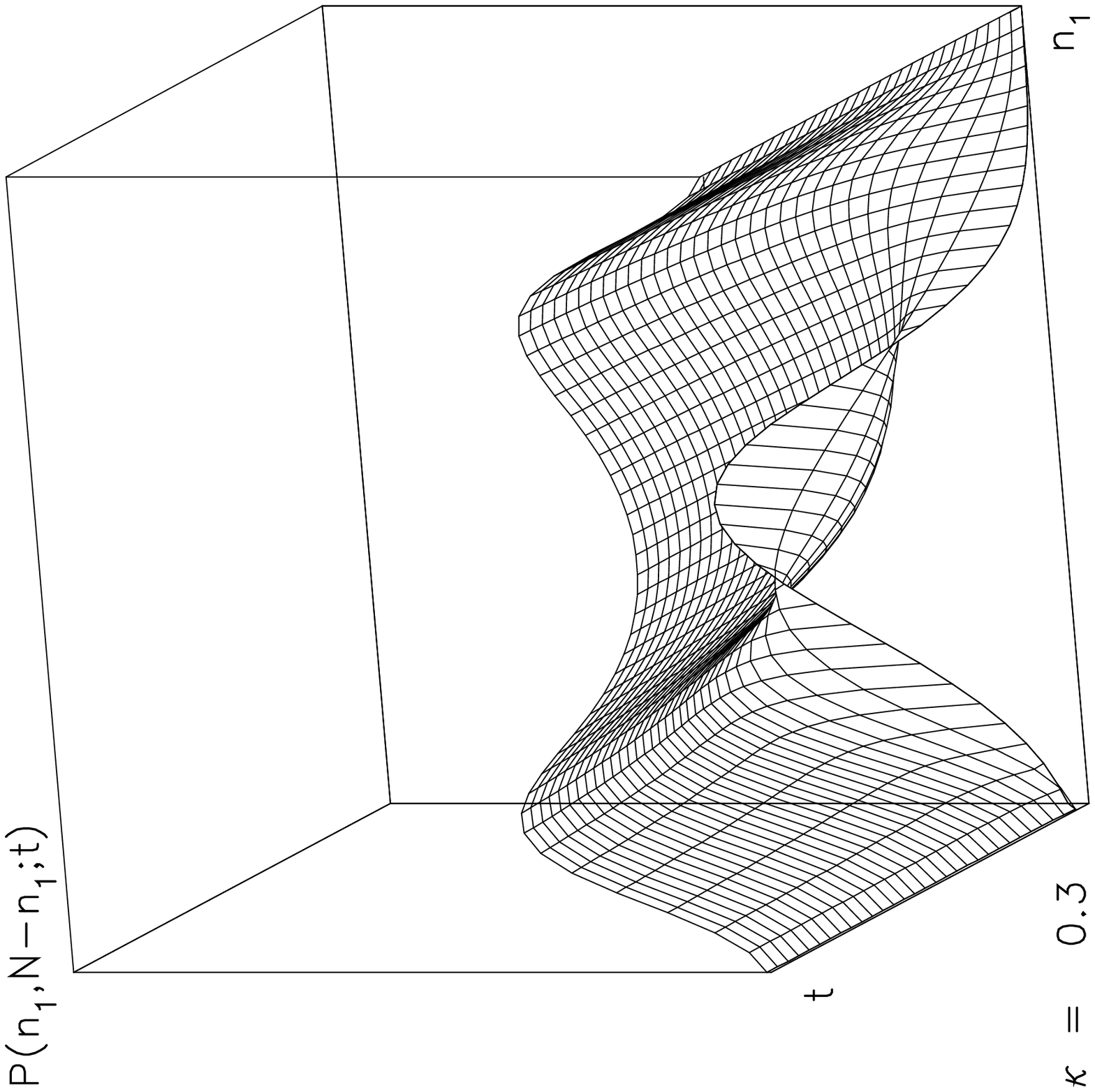}}}}
}
\parbox[b]{16cm}{\capt{Probability distribution $P(\vek{n},t)
\equiv P(n_1,N-n_1;t)$ of the (socio)configuration $\vek{n}$
for varying values of the control parameter $\kappa$. For $\kappa = 0$
a phase transition occurs: Whilst for $\kappa < 0$ both strategies
are used by about one half of the individuals, for $\kappa > 0$
very probably one of the strategies will be prefered after some time. 
That means, a behavioral convention
develops by social selforganization.\label{phasetr}}}
\end{figure}

We will now examine, if the behavioral model can explain this {\em break of
symmetry}. Figure \ref{phasetr} shows some computational 
results for $D(\vec{y},\vec{x};t) \equiv 2$ and $A=1$.
If 
\begin{equation}
 \kappa := 1 - 4W < 0 \, ,
\end{equation}
the configurational distribution is unimodal and symmetrical with respect to
$n_1 = N/2 = n_2$, i.e., both strategies will be chosen by about one
half of the individuals. 
At the {\em critical point} $\kappa = 0$
there appears a {\em phase transition}. This is indicated by the
broadness of the probability distribution 
$P(\vek{n},t) \equiv P(n_1,n_2;t) = P(n_1,N-n_1;t)$, 
which is due to {\em critical
fluctuations}.
For $\kappa > 0$ the 
configurational distribution becomes bimodal in the course of time, so that
one of the two equivalent strategies will very probably be chosen by
a majority of individuals. This can be interpreted as 
{\em selforganization of a behavioral convention}. 
Behavioral conventions often obtain a law-like character
after some time.

\section{The most probable strategy distribution}

In order to understand the phase transition more explicitly, we shall
in the following
consider the equations for the {\em most probable strategy distribution}
\begin{equation}
P(\vec{x},t) := \frac{\widehat{n}_{\weg{x}}(t)}{N} 
\end{equation}
with
\begin{equation} 
P(\vec{x},t) \ge 0 \qquad \mbox{and} \qquad \sum_{\weg{x}} P(\vec{x},t) = 1
\, ,
\end{equation}
where $\widehat{\vek{n}}(t)$ denotes the most probable
(socio)configuration. These equations are
approximately given by
\begin{equation}
 \frac{d}{dt} P(\vec{x},t) = 
 m_{\weg{x}}(\widehat{\vek{n}},t)  \, ,
\end{equation}
as can be seen by reformulating the master
equation (\ref{master}) in terms of a {\sc Langevin} equation 
\cite{Diss,Hel92}.
Here,
\begin{equation}
 m_{\weg{x}}(\widehat{\vek{n}},t) 
:= \sum_{\weg{x}'} \Big[ \overline{w}(\vec{x}|\vec{x}';
 \widehat{\vek{n}};t)P(\vec{x}',t)
 - \overline{w}(\vec{x}'|\vec{x};\widehat{\vek{n}};t) P(\vec{x},t) \Big] 
\label{mean1}
\end{equation}
are {\em drift coefficients}, and
\begin{equation}
\overline{w}(\vec{x}'|\vec{x};\widehat{\vek{n}};t) 
:=  w_1(\vec{x}'|\vec{x};\widehat{\vek{n}};t) 
+ \sum_{\weg{y}'}\sum_{\weg{y}}
w_2(\vec{x}',\vec{y}'|\vec{x},\vec{y};\widehat{\vek{n}};t)
\widehat{n}_{\weg{y}}
\label{mean2}
\end{equation}
have the meaning of {\em effective transition rates} \cite{Helb}. 
It turns out that the explicit equations for the 
most probable strategy distribution $P(\vec{x},t)$ 
are {\sc Boltzmann}-like equations:\alpheqn{Boltz}
\begin{eqnarray}
 \frac{d}{dt} P(\vec{x},t) &=& \sum_{\weg{x}'}
 \Big[ \widehat{w}_1(\vec{x}|\vec{x}';t) P(\vec{x}',t)
 - \widehat{w}_1(\vec{x}'|\vec{x};t)P(\vec{x},t) \Big] \label{term1} \\
 &+& \sum_{\weg{x}'}\sum_{\weg{y}}\sum_{\weg{y}'}
 \widehat{w}_2(\vec{x},\vec{y}'|\vec{x}',\vec{y};t)
 P(\vec{x}',t)P(\vec{y},t) \nonumber \\
 &-& \sum_{\weg{x}'}\sum_{\weg{y}}\sum_{\weg{y}'}
 \widehat{w}_2(\vec{x}',\vec{y}'|\vec{x},\vec{y};t)
 P(\vec{x},t)P(\vec{y},t) \label{term2}
\end{eqnarray}\reseteqn
with
\begin{eqnarray}
 \widehat{w}_1(\vec{x}'|\vec{x};t) &:= & 
 w_1(\vec{x}'|\vec{x};\widehat{\vek{n}}; t) \, , \\
 \widehat{w}_2(\vec{x}',\vec{y}'|\vec{x},\vec{y};t) &:= & N w_2
 (\vec{x}',\vec{y}'|\vec{x},\vec{y};\widehat{\vek{n}}; t) \, .
\end{eqnarray}
Obviously, the terms (\ref{term2}) are {\sc Boltzmann} (collision) terms
resulting from pair interactions, whereas the terms (\ref{term1})
are due to spontaneous transitions. 

\subsection{Oscillatory and chaotic behavior}\label{CHaoS}

\begin{figure}[htbp]
\parbox[b]{7.8cm}{
\epsfysize=7.8cm   
\centerline{\rotate[r]{\hbox{\epsffile[57 28 555
756]{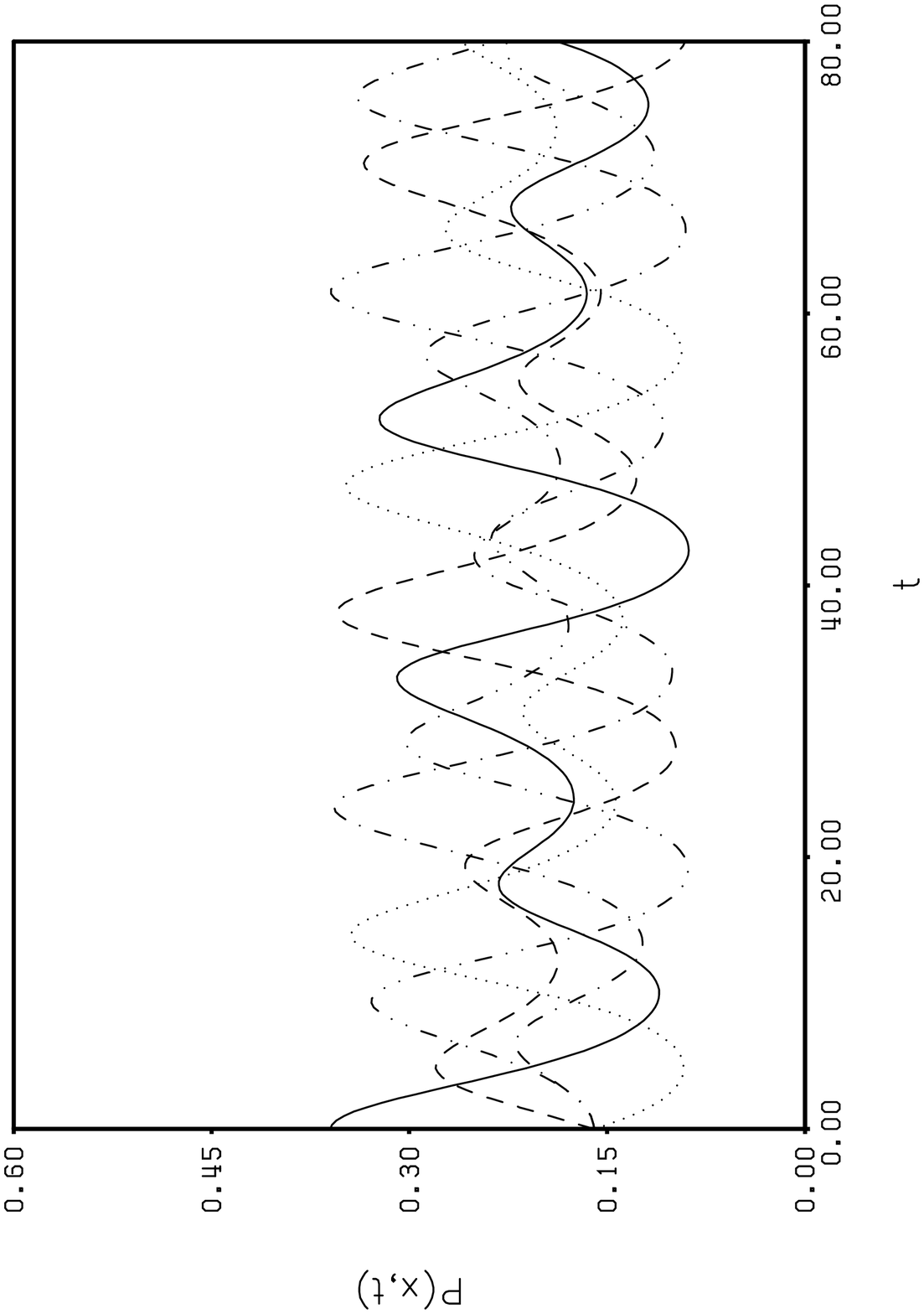}}}}
}\hfill
\parbox[b]{7.8cm}{
\epsfysize=7.4cm                   
\centerline{\rotate[r]{\hbox{\epsffile[57 40 555 613]
{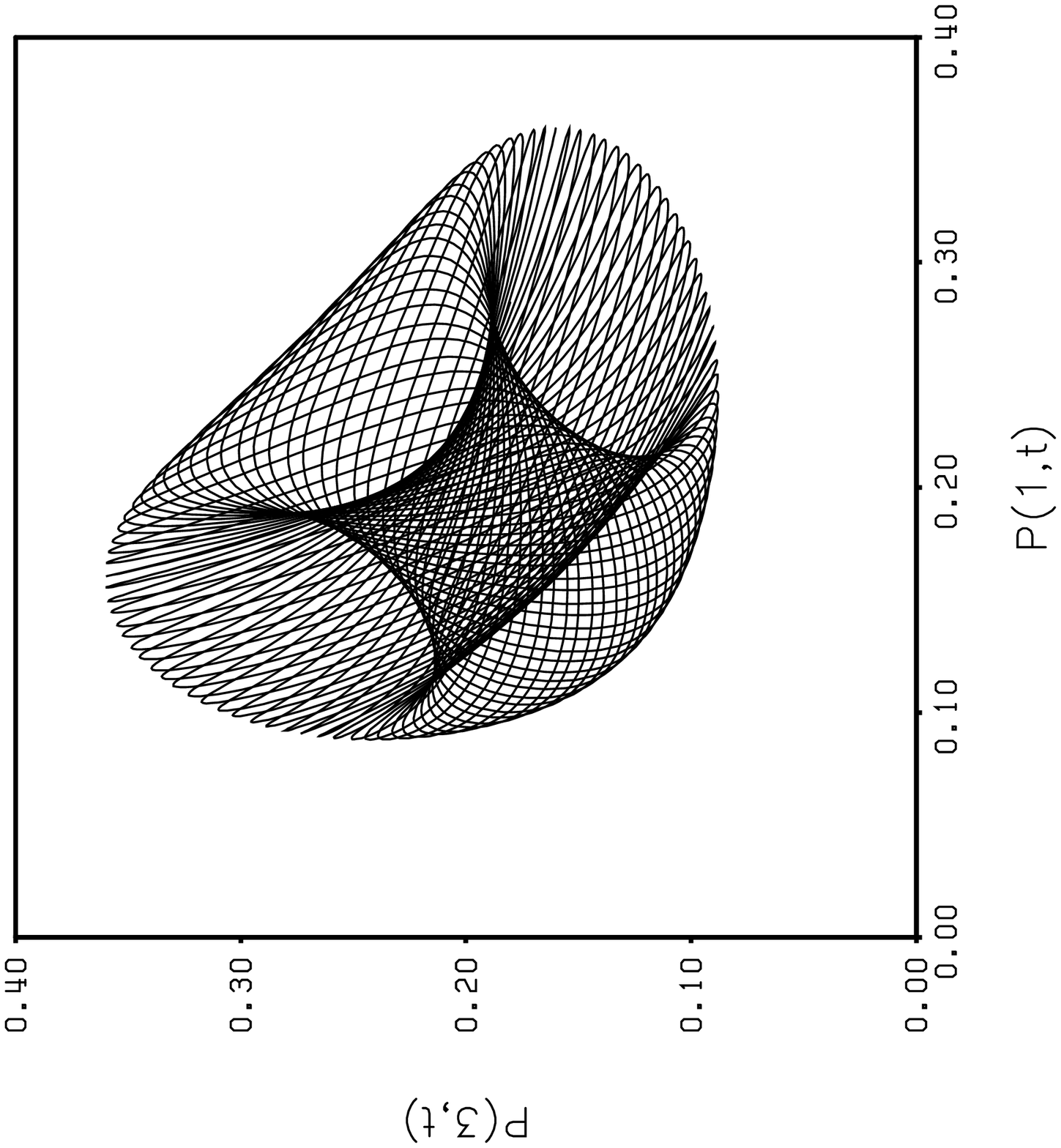}}}}
}
\parbox[b]{16cm}{\capt{Oscillations are one possible effect of imitative 
processes. For $S=5$ different strategies, the oscillatory changes look
quite irregular without a short-term periodicity. The
corresponding phase portrait has the shape of a torus, which indicates 
a long-term periodicity by the closeness of the curve.
\label{oszi}}
}
\end{figure}

\begin{figure}[htbp]
\parbox[b]{7.8cm}{
\epsfysize=7.8cm   
\centerline{\rotate[r]{\hbox{\epsffile[57 28 555
756]{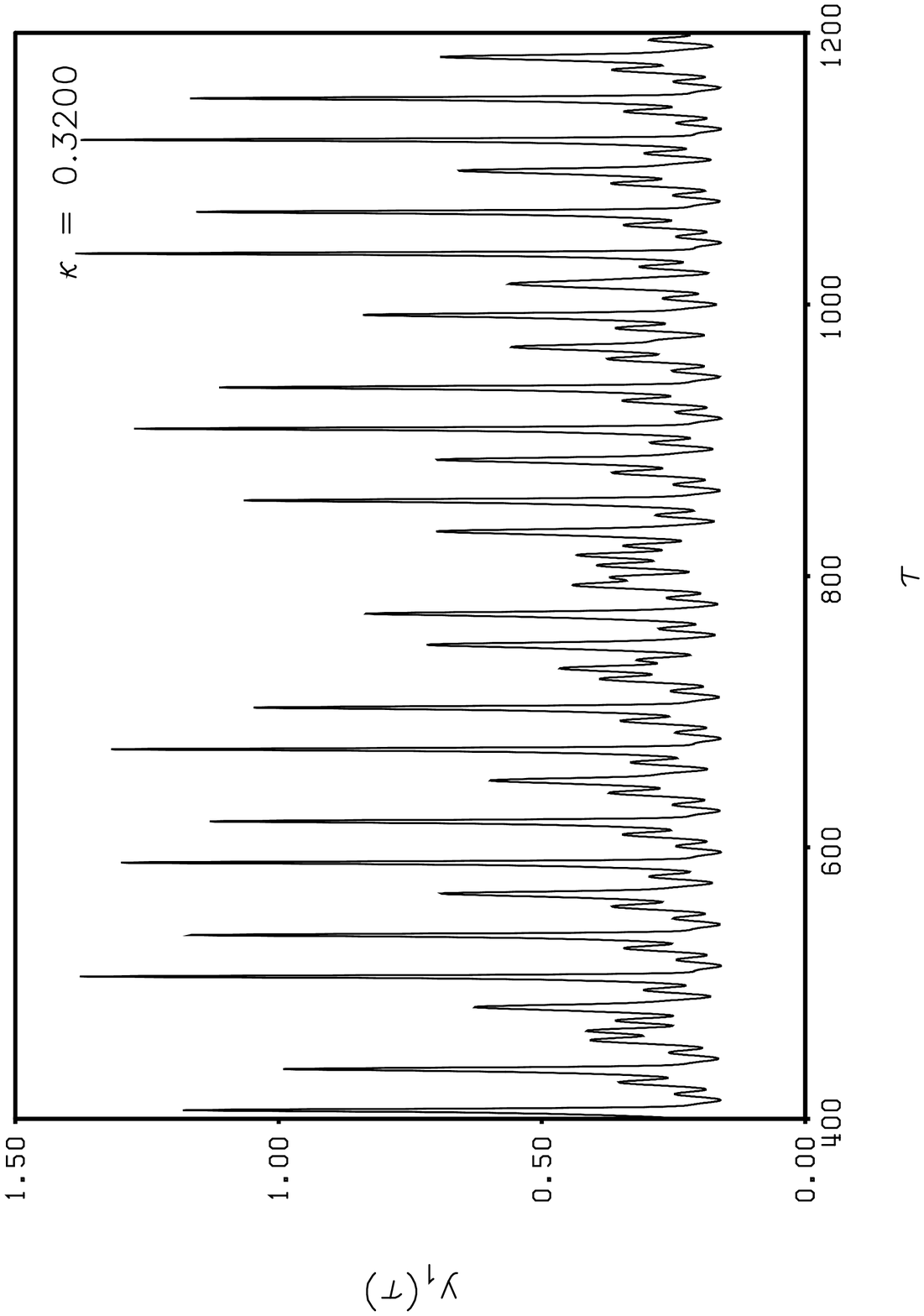}}}}
}\hfill
\parbox[b]{7.8cm}{
\epsfysize=7.8cm                          
\centerline{\rotate[r]{\hbox{\epsffile[57 28 555 
756]{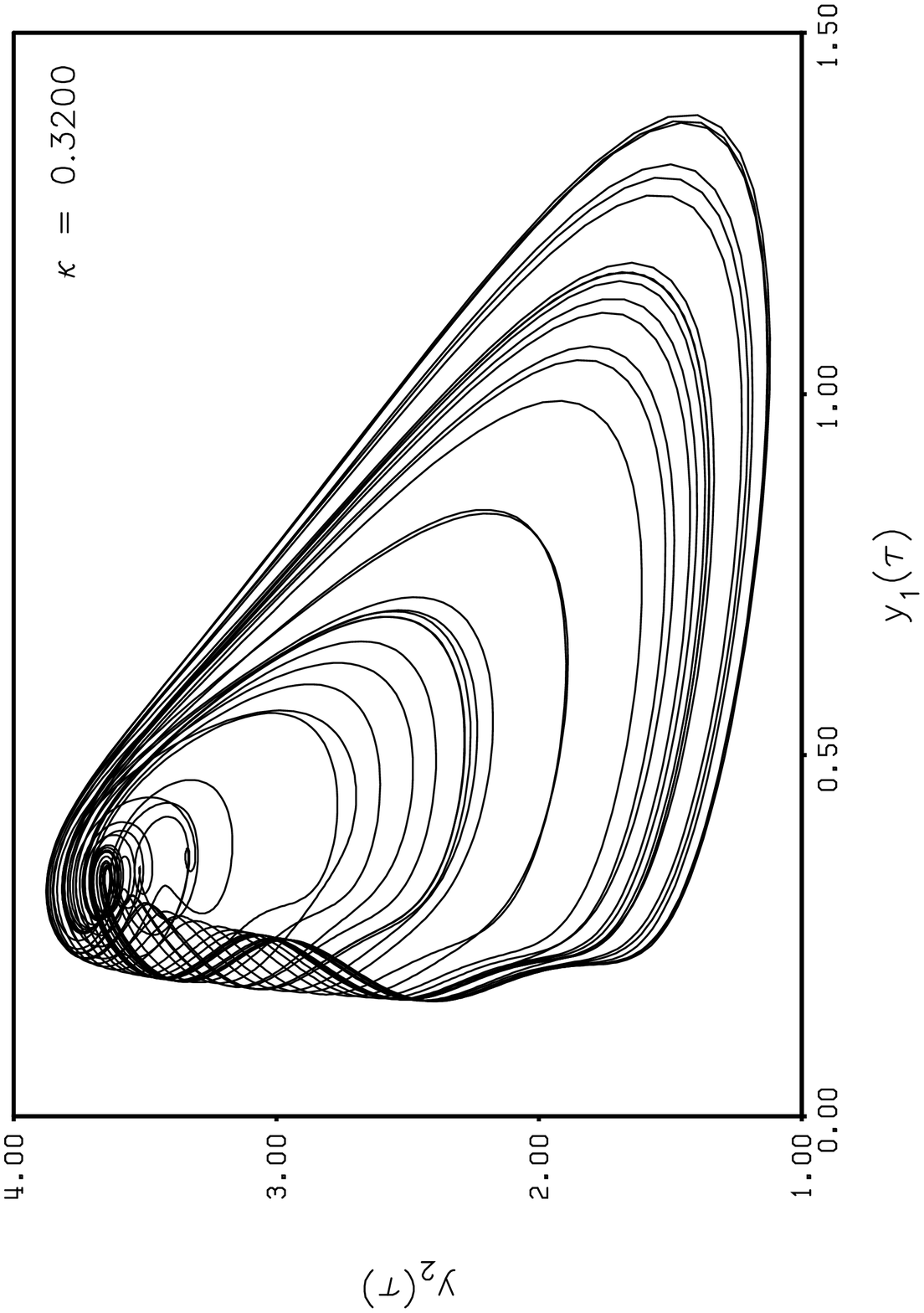}}}}
}
\parbox[b]{7.8cm}{
\epsfysize=7.8cm   
\centerline{\rotate[r]{\hbox{\epsffile[14 28 555
756]{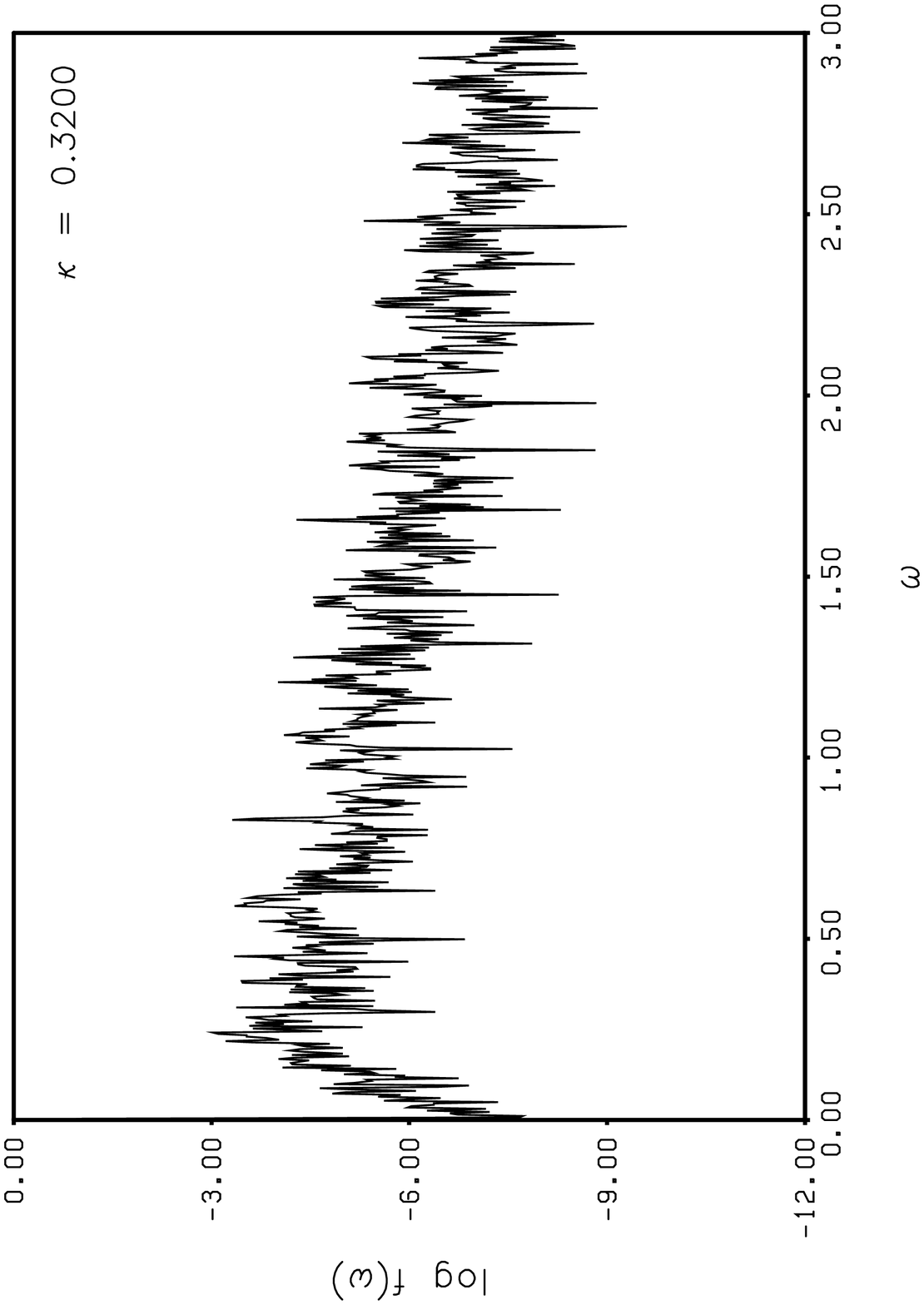}}}}
}\hfill
\parbox[b]{7.8cm}{
\epsfysize=7.8cm                          
\centerline{\rotate[r]{\hbox{\epsffile[14 28 555 
756]{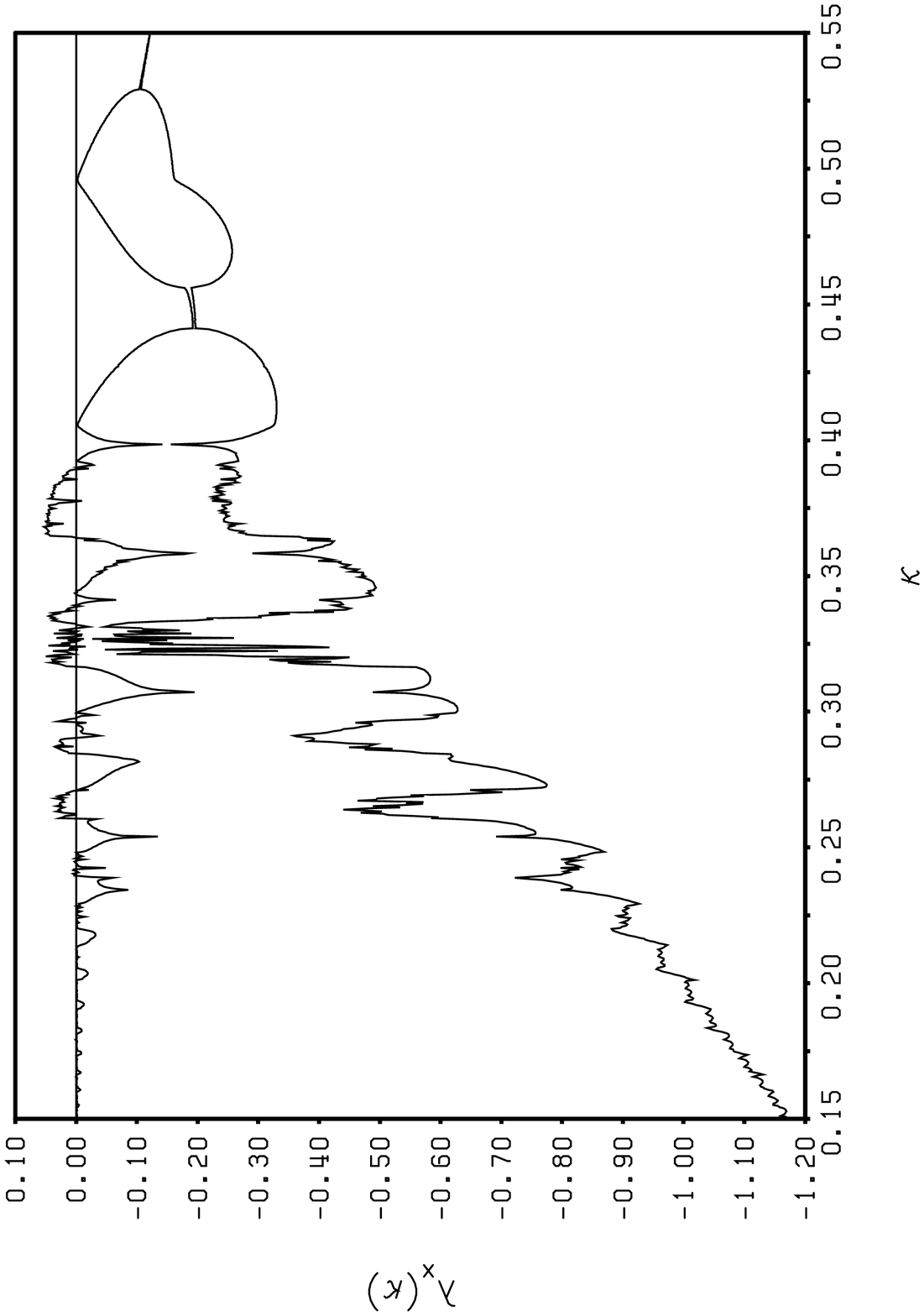}}}}
}
\parbox[b]{16cm}{\capt{Temporal development, phase portrait,
power spectrum $f(\omega)$ and greatest {\sc Lyapunov} exponents 
$\lambda_x(\kappa)$ of the scaled variables 
$y_x(\tau) = \mu_x P(x,\nu t)$ for special
{\sc Boltzmann} equations that produces chaos.\label{chaot}}}
\end{figure}

For $w_1(\vec{x}'|\vec{x};\widehat{\vek{n}};t) \equiv 0$, 
$w_2(\vec{x}',\vec{y}'|\vec{x},
\vec{y};\widehat{\vek{n}};t) \equiv w_2(\vec{x}',\vec{y}'|\vec{x},
\vec{y})$, and
\begin{equation}
 \sum_{\weg{x}'}\sum_{\weg{y}'} w_2(\vec{x},\vec{y}|\vec{x}',\vec{y}')
= \sum_{\weg{x}'}\sum_{\weg{y}'} w_2(\vec{x}',\vec{y}'|\vec{x},\vec{y}) 
\label{bedin}
\end{equation}
equation (\ref{Boltz}) obeys the famous {\sc Boltzmann} 
$H$-theorem \cite{Diss}
\begin{equation}
 \frac{dH}{dt} \le 0 \qquad \mbox{with} \qquad
 H(t) := \sum_{\weg{x}} P(\vec{x},t) \ln P(\vec{x},t) \, .
\end{equation}
According to the $H$-theorem $P(\vec{x},t)$ approaches a stationary
solution $P_0(\vec{x})$ in the course of time. For example, in a dilute
gase the velocity distribution approaches a {\sc Maxwell} distribution.
However, for social processes the relation (\ref{bedin})
may be invalid (since there
are no collisional invariants). As a consequence, the corresponding
{\sc Boltzmann}-like equations can show oscillatory 
or even chaotic solutions \cite{Diss,Hel92} (see figures \ref{oszi}
and \ref{chaot}). 
\par
For example, the special {\sc Boltzmann} equations
\begin{equation}
 \frac{d}{dt} P(\vec{x},t) = \nu P(\vec{x},t) \Big[ P(\vec{x} - 1,t)
 - P(\vec{x} + 1,t) \Big] 
\qquad \mbox{with} \qquad  \vec{x} \equiv \vec{x}\mbox{ mod } S 
\end{equation}
display {\em nonlinear oscillations} (see fig. \ref{oszi}): 
A linear stability analysis around
the stationary point $\vek{P}_0 := (1/S,\dots,1/S)^{\rm tr}$ shows that the
eigenvalues of the corresponding {\sc Jacobi}an matrix are
purely imaginary \cite{Diss,He92}. 
Due to the relations
\begin{equation}
\sum_{\weg{x}=1}^S P(\vec{x},t) = 1
\qquad \mbox{and} \qquad 
 \prod_{\weg{x}=1}^S P(\vec{x},t) = \mbox{\it const.}
\end{equation}
the trajectory $\vek{P}(t) \equiv
(P(1,t),\dots,P(S,t))^{\rm tr}$ moves on a $(S-2)$-dimensional
hypersurface. For $S=3$ strategies
the shape of the resulting cycles can be calculated
explicitly. It is given by
\begin{equation}
 P(2,t) = \frac{1 - P(1,t)}{2} \pm \sqrt{ \left[ \frac{1-P(1,t)}{2} \right]^2
 - \frac{C}{P(1,t)} }
\end{equation}
with
\begin{equation}
 P(3,t) = 1 - P(1,t) - P(2,t) \qquad \mbox{and} \qquad
 C := P(1,t_0)P(2,t_0)P(3,t_0) \, .
\end{equation}
{\sc Boltzmann} equations with chaotic solutions are, for example, given by
the following interaction schemes:
\begin{eqnarray}
4,9 & \stackrel{k_1}{\longrightarrow} & 1,9 \, , \nonumber \\ 
1,9 & \stackrel{k'_1}{\longrightarrow} & 4,9 \, , \nonumber \\ 
1,9 & \stackrel{k_2}{\longrightarrow} & 2,9 \, , \nonumber \\ 
4,6 & \stackrel{k_3}{\longrightarrow} & 1,6 \, , \nonumber \\ 
1,1 & \stackrel{k_4}{\longrightarrow} & 3,3 \, , \nonumber \\ 
3,9 & \stackrel{k'_4}{\longrightarrow} & 1,9 \, , \nonumber \\ 
2,3 & \stackrel{k^{\prime\prime}_4}{\longrightarrow} & 1,3 \, , \nonumber \\ 
8,6 & \stackrel{k_5}{\longrightarrow} & 6,6 \, , \nonumber \\ 
6,7 & \stackrel{k'_5}{\longrightarrow} & 7,7 \, , \nonumber \\ 
7,8 & \stackrel{k^{\prime\prime}_5}{\longrightarrow} & 8,8 \, , \nonumber \\ 
4,4 & \stackrel{k_6}{\longrightarrow} & 4,5 \, , \nonumber \\ 
5,4 & \stackrel{k'_6}{\longrightarrow} & 4,4 \, , 
\end{eqnarray}
where $k_l$ denote the interaction rates $\widehat{w}_2(\vec{x}',\vec{y}'|
\vec{x},\vec{y})$ of the pair interactions
\begin{equation}
\qquad  \vec{x} , \vec{y} \; 
\stackrel{k_l}{\longrightarrow} \;  \vec{x}', \vec{y}' 
\, . \qquad
\end{equation}
Using the abbreviations
\begin{equation}
\begin{array}{rclrclrcl}\displaystyle
\alpha &:=& \sqrt{k'_1k'_4/(2k_4k_4^{\prime\prime})} \, , \quad &
\beta  &:=& k'_1/k_4^{\prime\prime}  \, , \quad &
\gamma &:=& \alpha k_6/k'_6 \, , \\
\tau(t)&:=& k'_1P(9,0)t \, , \quad &
a &:=& k_1/k'_1 \, , \quad &
b &:=& k_2/k'_1 \, , \\ 
c &:=& \alpha k_5/k'_1 \, , \quad &
c'&:=& \alpha k'_5/k'_1 \, , \quad &
c^{\prime\prime}&:=& \alpha k_5^{\prime\prime}/k'_1 \, , \\
d &:=& \alpha k_6/k'_1 \, , \quad &
e &:=& k_6/k'_6 \, , \quad &
\kappa &:=& \alpha k_3/k'_1 \, , \\
\epsilon &:=& 2 \alpha k_4/k'_1 \, , \quad &
\epsilon'&:=& k'_1/(k_4^{\prime\prime}\alpha) \, . \quad &
 & & 
\end{array}
\end{equation}
and the scaled variables $y_x(\tau)$ according to
\begin{equation}
 P(x,t) =: y_x(\tau) P(9,0) \cdot \left\{
\begin{array}{ll}
\alpha & \mbox{if } x \in \{1,2,4,6,7,8\} \\
\beta & \mbox{if } x = 3 \\
\gamma & \mbox{if } x = 5 \\
1 & \mbox{if } x = 9 \, ,
\end{array} \right.
\end{equation}
the corresponding {\sc Boltzmann} equations are:
\begin{eqnarray*}
 \frac{d}{d\tau} y_1(\tau)&=& a y_4(\tau)y_9(\tau) - (b +1)y_1(\tau)y_9(\tau) 
+ \kappa y_4(\tau)y_6(\tau) \nonumber \\
&-& \epsilon \Big[ [y_1(\tau)]^2 - y_3(\tau)y_9(\tau)\Big] + y_2(\tau)y_3(\tau) 
\, , \nonumber \\
\frac{d}{d\tau}y_2(\tau) &=& b y_1(\tau)y_9(\tau) - y_2(\tau)y_3(\tau) 
\, , \nonumber \\
{\epsilon'}\frac{d}{d\tau}y_3(\tau) &=& \epsilon 
\Big[ [y_1(\tau)]^2 - y_3(\tau)y_9(\tau) \Big]
\, , \nonumber \\
\end{eqnarray*}
\begin{eqnarray}
\frac{d}{d\tau}y_4(\tau) &=& - a y_4(\tau)y_9(\tau) + y_1(\tau)y_9(\tau) 
 - \kappa y_4(\tau)y_6(\tau) \nonumber \\
&-& d y_4(\tau)\Big[y_4(\tau) - y_5(\tau)\Big] \, , \nonumber \\
 e \frac{d}{d\tau}y_5(\tau) &=& d
 y_4(\tau)\Big[y_4(\tau) - y_5(\tau)\Big] \, , \nonumber \\
\frac{d}{d\tau} y_6(\tau) &=& y_6(\tau)
\Big[ c y_8(\tau) -  c' y_7(\tau)\Big] \, , \nonumber \\
\frac{d}{d\tau} y_7(\tau) &=& y_7(\tau)
\Big[ c' y_6(\tau) - c^{\prime\prime}y_8(\tau)\Big] \, , \nonumber \\
\frac{d}{d\tau} y_8(\tau) &=& y_8(\tau)
\Big[ c^{\prime\prime}y_7(\tau)- c y_6(\tau)\Big] \, , \nonumber \\
\frac{d}{d\tau} y_9(\tau) &=& 0 \, .
\end{eqnarray}
For certain sets of parameters these equations have {\em chaotic} solutions.
Especially, for the parameters
\begin{equation}
\begin{array}{rclrclrcl}
a &:=& 0 \, , \quad &
b &:=& 1.2 \, , \quad & 
& & \\
c &:=& 0.46 \, , \quad &
c'&:=& 0.46 \, , \quad &
c^{\prime\prime}&:=&0.46 \, , \\
d &:=& 100 \, , \quad &
e &:=& 10000 \, , \quad &
\kappa &=& \mbox{\it varying} \, , \\
\epsilon &:=& 0.01 \, , \quad &
\epsilon'&:=& 0.0001 \quad &
 & & 
\end{array}
\label{consts}
\end{equation}
and the initial conditions
\begin{equation}
\begin{array}{rclrcl}
y_1(0) &:=& 0.6 \, ,  \qquad &
y_2(0) &:=& 1.8 \, , \\
y_3(0) &:=& 0.36 \, , \qquad &
y_4(0) &:=& 1 \, , \\
y_5(0) &:=& 1 \, , \qquad &
y_6(0) &:=& 1.12 \, , \\
y_7(0) &:=& 1 + 0.12 \sin\left(2\pi/3 \right) \, , \qquad &
y_8(0) &:=& 1 + 0.12 \sin\left(4\pi/3 \right) 
\end{array}
\end{equation}
several {\em period doubling sequences} are found, if $\kappa \in [0.15,0.55]$
is varied (see \cite{Diss} for a more detailled discussion). 
Figure \ref{chaot} shows computational results for $\kappa = 0.32$.

\section{The game dynamical equations}
              
The socalled {\em game dynamical equations}, which are
used for the description of cooperation and competition processes,
are also a special case of the 
{\sc Boltzmann}-like equations. This is shown in the
following:
If we again restrict our considerations to imitative 
processes (see (\ref{imita})), 
the special {\sc Boltzmann}-like equations
\begin{eqnarray}
 \frac{d}{dt} P(\vec{x},t) &=& \sum_{\weg{x}'} 
 \Big[ \widehat{w}_1(\vec{x}|\vec{x}';t)
 P(\vec{x}',t) 
- \widehat{w}_1(\vec{x}'|\vec{x};t) P(\vec{x},t) \Big] \nonumber \\
 &+& P(\vec{x},t) \sum_{\weg{x}'} \Big[ \widehat{w}_2^*(\vec{x}|\vec{x}';t)
 - \widehat{w}_2^*(\vec{x}'|\vec{x};t) \Big] P(\vec{x}',t) 
\label{special}
\end{eqnarray}
result with
\begin{equation}
 \widehat{w}_2^*(\vec{x}'|\vec{x};t) := N w_2^*(\vec{x}'|\vec{x};
\widehat{\vek{n}};t) 
\, .
\end{equation}
Inserting the multinomial logit ansatz (\ref{mult}) and 
using a suitable {\sc Taylor} approximation leads to the
{\em game dynamical equations}\alpheqn{Game}
\begin{eqnarray}
\frac{d}{dt} P(\vec{x},t) 
&=& \sum_{\weg{x}'} \Big[ \widehat{w}_1(\vec{x}|\vec{x}';t)P(\vec{x}',t) 
- \widehat{w}_1(\vec{x}'|\vec{x};t)P(\vec{x},t) \Big] \label{mutation} \\
&+& P(\vec{x},t) \Big[ \widehat{E}(\vec{x},t) 
- \langle \widehat{E} \rangle \Big] \, ,
\label{selection}
\end{eqnarray}\reseteqn
where 
\begin{equation}
 \widehat{E}(\vec{x},t) := E(\vec{x};\widehat{\vek{n}};t)
= \sum_{\weg{x}'} A_{\weg{x}\weg{x}'} 
 \frac{\widehat{n}_{\weg{x}'}(t)}{N} 
\end{equation}
is the {\em expected success} of strategy $\vec{x}$, and
\begin{equation}
 \langle \widehat{E} \rangle := \sum_{\weg{x}'} 
 \widehat{E}(\vec{x}',t) P(\vec{x}',t) 
\end{equation}
is the {\em mean expected success}.
Since (\ref{selection}) can be 
understood as effect of a {\em selection}
(of strategies with an expected success that exceeds the
average $\langle \widehat{E} \rangle $) and
(\ref{mutation}) can be interpreted as effect of spontaneous strategy changes
(e.g. due to accidental {\em mutations}) the game dynamical equations are
also known as {\em selection mutation equations} \cite{HoSi88,EbFe82}.
The mutation term can be used for the description
of {\em trial and error}. 
\par
As an example, we shall again examine the case of two equivalent strategies.
The game dynamical equations (\ref{Game}) corresponding
to (\ref{pay}), (\ref{fluct}) have, then, the explicit form
\begin{equation}
 \frac{d}{dt} P(\vec{x},t) = -2\left( P(\vec{x},t) - \frac{1}{2} \right) 
\Big[ W + AP(\vec{x},t) \Big( P(\vec{x},t) - 1 \Big) \Big] \, . 
\label{concr}
\end{equation}
According to (\ref{concr}), $P(\vec{x}) = 1/2$
is a stationary solution. This solution is stable for
\begin{equation}
 \kappa := 1 - \frac{4W}{A} < 0 \, ,
\end{equation}
i.e., if spontaneous strategy changes are dominating and, therefore, 
prevent a selforganization process.
\par
At the {\em critical point} $\kappa = 0$ 
there appears a {\em break of symmetry}: 
For $\kappa > 0$ the stationary solution $P(\vec{x}) = 1/2$ is unstable, and
the game dynamical equations (\ref{concr}) can be rewritten in the form
\begin{equation}
 \frac{d}{dt} P(\vec{x},t) = -2 \left( P(\vec{x},t) - \frac{1}{2} \right) 
\left( P(\vec{x},t) - \frac{1 + \sqrt{\kappa}}{2} \right) 
\left( P(\vec{x},t) - \frac{1-\sqrt{\kappa}}{2} \right) \, .
\end{equation}
That means, for $\kappa > 0$ we have two additional stationary solutions 
$P(\vec{x}) = (1+\sqrt{\kappa})/2$ and 
$P(\vec{x}) = (1-\sqrt{\kappa})/2$, which
are stable. Depending on initial fluctuations,
one strategy
will win a majority of $100\cdot\sqrt{\kappa}$ 
percent. This majority is the greater,
the smaller the rate $W$ of spontaneous strategy changes is.
\par
The game dynamical equations (including generalizations and other derivations)
are more explicitly discussed in \cite{Diss,Hel92}.

\section{Summary and Conclusions}

The master equation and {\sc Boltzmann}-like equations have shown to be 
suitable for the quantitative description of
behavioral changes and social processes.
In the models developed spontaneous strategy changes 
and behavioral changes due to pair interactions have been taken into
account.
Three kinds of pair interactions have been distinguished:
imitative, avoidance and compromising processes.
The game dynamical equations result for a special case of imitative processes.
They can be interpreted as equations for the most probable behavioral
distribution and allow the description of social selforganization
of behavioral conventions.
{\small\itemsep-1mm
\subsection*{Acknowledgements}
The author is grateful to Prof. Dr. W. Weidlich and
PD Dr. G. Haag for inspiring discussions. He also wants to thank
M. Schanz, who provided the FORTRAN routines 
for the calculation of the {\sc Lyapunov} exponents.

}

\begin{thebibliography}{99}                         
%
\bibitem{Zwan} R. Zwanzig.                           
On the identity of three generalized master equations.
{\it Physica} {\bf 30}, 1109--1123, 1964.
%
\bibitem{Opp} I. Oppenheim, K. E. Schuler and G. H. Weiss, eds.
{\it Stochastic Processes in Chemical Physics: The Master Equation}.
MIT Press, Cambridge, Mass., 1977.
%
\bibitem{Hak} H. Haken. {\it Laser Theory}. Springer, Berlin, 1984.
%
\bibitem{Arn} L. Arnold and R. Lefever, eds. {\it Stochastic Nonlinear Systems
in Physics, Chemistry and Biology}. 
Springer, Berlin, 1981.
%
\bibitem{WeHa83} W. Weidlich \& G. Haag.
{\it Concepts and Models of a Quantitative Sociology. The Dynamics of
Interacting Populations}.
Springer, Berlin, 1983.
%
\bibitem{We91} W. Weidlich.
Physics and social science---{T}he approach of synergetics.
{\it Physics Reports} {\bf 204}, 1--163, 1991.
%
\bibitem{Weid1} W. Weidlich. 
The use of statistical models in sociology.
{\it Collective Phenomena} {\bf 1}, 51-59, 1972.
%
\bibitem{Weid2} W. Weidlich \& G. Haag, eds. {\it Interregional Migration}.
Springer, Berlin, 1988.
%
\bibitem{WeHa87} W. Weidlich \& G. Haag.
A dynamic phase transition model for spatial agglomeration processes.
{\it Journal of Regional Science} {\bf 27}(4), 529--569, 1987.
%
\bibitem{WeMu90} W. Weidlich \& M. Munz.
Settlement formation, Part I: A dynamic theory.
{\it Annals of Regional Science} {\bf 24}, 83--106, 1990.
%
\bibitem{Bo64} L. Boltzmann.
{\it Lectures on Gas Theory}.
University of California, Berkeley, 1964.
%
\bibitem{Br82} R. Brdi\v{c}ka.
{\it {G}rundlagen der physikalischen {C}hemie}, Kap.~9: {R}eaktionskinetik.
Deutscher Verlag der Wissenschaften, Berlin, 15. Auflage, 1982.
%
\bibitem{Pea24} R. Pearl.
{\it Studies in Human Biology}.
Williams \& Wilkins, Baltimore, 1924.
%
\bibitem{Ve45} P.~F. Verhulst.
{\it Nuov. Mem. Acad. Roy. Bruxelles} {\bf 18}, 1, 1845.
%
%
\bibitem{Zi46} G.~K. Zipf.
The {P}1{P}2/{D} hypothesis on the intercity movement of persons.
{\it American Sociological Review} {\bf 11}, 677--686, 1946.
%
\bibitem{Diss} D. Helbing.
{\it Stochastische Methoden, nichtlineare Dynamik und quantitative Modelle
sozialer Prozesse}.
Dissertation, Universit{\"a}t Stuttgart, 1992.
%
\bibitem{Compl} D. Helbing. 
A fluid dynamic model for the movement 
of pedestrians. Submitted to {\it Complex Systems}.
%
\bibitem{He92} D. Helbing.
A mathematical model for attitude formation by pair interactions.
{\it Behavioral Science} {\bf 37}, 190--214, 1992.
%
\bibitem{Co84} A.~M. Colman.
{\it Game Theory and Experimental Games}.
Pergamon Press, Oxford, 1984.
%
\bibitem{Ra90} A. Rapoport.
{\it Experimental Studies of Interactive Decisions}.
Kluwer Academic, Dordrecht, 1990.
%
\bibitem{Ax84} R. Axelrod.
{\it The Evolution of Cooperation}.
Basic Books, New York, 1984.
%
\bibitem{Mue90} U. Mueller, Hrsg.
{\it {E}volution und {S}pieltheorie}.
Oldenbourg, M{\"u}nchen, 1990.
%
\bibitem{NeuMo44} J. von~Neumann \& O. Morgenstern.
{\it Theory of Games and Economic Behavior}.
Princeton University Press, Princeton, 1944.
%
\bibitem{Ei71} M. Eigen.
The selforganization of matter and the evolution of biological macromolecules.
{\it Naturwissenschaften} {\bf 58}, 465, 1971.
%
\bibitem{Fi30} R.~A. Fisher.
{\it The Genetical Theory of Natural Selection}.
Oxford University Press, Oxford, 1930.
%
\bibitem{EiSchu79} M. Eigen \& P. Schuster.
{\it The Hypercycle}.
Springer, Berlin, 1979.
%
\bibitem{FeEb89} R. Feistel \& W. Ebeling.
{\it Evolution of Complex Systems}.
Kluwer Academic, Dordrecht, 1989.
%
%
\bibitem{Lo56} A.~J. Lotka.
{\it Elements of Mathematical Biology}.
Dover, New York, 1956.
%
\bibitem{Vo31} V. Volterra.
{\it Le\c{c}ons sur la th\'{e}orie math\'{e}matique de la lutte pour la vie}.
Gauthier-Villars, Paris, 1931.
%
\bibitem{GoMaMo71} N.~S. Goel, S.~C. Maitra  \& E.~W. Montroll.
On the {V}olterra and other nonlinear models of interacting populations.
{\it Reviews of Modern Physics} {\bf 43}(2), 231--276, 1971.
%
\bibitem{HoSi88} J. Hofbauer \& K. Sigmund.
{\it The Theory of Evolution and Dynamical Systems}.
Cambridge University Press, Cambridge, 1988.
%
\bibitem{Helb} D. Helbing.
Interrelations between stochastic equations for systems with pair interactions.
{\it Physica A} {\bf 181}, 29--52, 1992.
%
\bibitem{Hel92} D. Helbing.
A mathematical model for behavioral changes by pair interactions and its
relation to game theory.
{\it Angewandte Sozialforschung} {\bf 17} (3/4), 179--194, 1992.
%
\bibitem{DoFa75} Th.~A. Domencich \& D. McFadden.
{\it Urban Travel Demand. A Behavioral Analysis}, pp.~61--69.
North-Holland, Amsterdam, 1975.
%
\bibitem{Lu59} R.~D. Luce.
{\it Individual Choice Behavior}, chap.~2.A: {F}echner's Problem.
Wiley, New York, 1959.
%
\bibitem{He91} D. Helbing.
A mathematical model for the behavior of pedestrians.
{\it Behavioral Science} {\bf 36}, 298--310, 1991.
%
\bibitem{EbFe82} W. Ebeling \& R. Feistel.
{\it Physik der Selbstorganisation und Evolution}.
Akademie-Verlag, Berlin, 1982.
\end{thebibliography}
\end{document}